\documentclass[conference,a4paper]{APSIPA2026}
\usepackage{amsmath}
\usepackage{amssymb}
\usepackage{graphicx}
\usepackage{stfloats}
\usepackage{cuted}
\usepackage{multirow}
\usepackage{threeparttable}
\usepackage{booktabs}
\usepackage{array}
\usepackage[colorlinks=true,citecolor=blue,linkcolor=blue,urlcolor=blue]{hyperref}
\usepackage[backend=biber,style=ieee,sorting=none]{biblatex}
\usepackage{geometry}
\usepackage{fancyhdr}

\fancypagestyle{firststyle}{
  \fancyhf{}
  \fancyhead[C]{2026 Asia Pacific Signal and Information Processing Association Annual Summit and Conference (APSIPA ASC)}
}

\newcommand{\xwm}{x_{\mathrm{wm}}}
\newcommand{\xrec}{x_{\mathrm{rec}}}
\newcommand{\xref}{x_{\mathrm{ref}}}

\newcommand{\mhat}{\hat{m}}
\newcommand{\Det}{\mathrm{Det}}
\newcommand{\Bit}{\mathrm{Bit}}

\begin{document}

\title{NeuMark: Neural Codec Resynthesis-Robust Audio Watermarking in the Codec Latent Space}

\author{
\authorblockN{
Annan Wu,
Wen-Chin Huang,
Tomoki Toda
}

\authorblockA{
Nagoya University, Japan \\
E-mail: wu.annan@g.sp.m.is.nagoya-u.ac.jp \\
E-mail: wen.chinhuang@g.sp.m.is.nagoya-u.ac.jp \\
E-mail: tomoki@icts.nagoya-u.ac.jp}
}

\maketitle
\thispagestyle{firststyle}
\pagestyle{empty}

\begin{abstract}
Audio watermarking is increasingly important for tracing generated speech.  Several audio watermarking methods have been proposed to embed the watermark in various domains, such as waveform, timbre feature, or latent representations, for making the embedded watermark robust against traditional digital signal processing (DSP) attacks.  On the other hand, modern neural codecs introduce a different threat from DSP attacks: they resynthesize speech through quantized acoustic representations and can remove the embedded watermark evidence that is not aligned with codec-preserved structure.  In this paper, we propose NeuMark, a codec-latent audio watermarking framework that embeds watermark evidence into SpeechTokenizer acoustic tokens to address this resynthesis threat.  NeuMark uses cross-attention to inject a 16-bit message across residual vector quantization (RVQ) layers, distributing the watermark over codec-aligned latent structure.  Experimental results show that NeuMark substantially improves robustness under neural-codec resynthesis while supporting both watermark detection and message recovery.  We also analyze the trade-off between reconstruction-referenced transparency and original-referenced robustness.
\end{abstract}


\section{Introduction}
Recent progress in text-to-speech (TTS), voice conversion, and neural audio codec language models, such as VALL-E \cite{wang2023valle} and AutoVC \cite{qian2019autovc}, has made synthetic speech increasingly natural and easy to personalize.  This progress also raises practical risks: impersonation, unauthorized voice cloning, and redistribution of generated speech without provenance.  Passive speech deepfake detectors can be useful, but they often depend on artifacts of specific generators and can degrade as synthesis models improve \cite{jung2022aasist,zhu2024slim}.  Audio watermarking addresses this provenance problem by embedding traceable evidence into speech: given speech and, optionally, a source- or user-specific message to be embedded, a watermarking system outputs perceptually similar watermarked speech.  The embedded evidence can later be used to verify the watermark and recover the message, enabling released speech to be traced to its generation pipeline or authorized source.  For this tracing to remain useful in practice, the watermark must survive benign and adversarial distortions while remaining inaudible.

Previous audio watermarking methods have embedded watermark evidence in several domains.  Waveform-domain methods such as WavMark \cite{chen2023wavmark} and AudioSeal \cite{sanroman2024audioseal} directly modify the generated waveform, while Timbre Watermarking \cite{liu2023timbrewm} embeds information in timbre-related representations.  Latent-side or generation-aware methods such as TraceableSpeech \cite{zhou2024traceablespeech} and VoiceMark \cite{li2025voicemark} inject watermark information before decoding or during synthesis.  These methods have shown robustness to many DSP attacks, such as filtering, noise, resampling, and volume changes.

However, robustness to DSP attacks does not necessarily imply robustness to neural-codec resynthesis.  Modern neural codecs such as EnCodec \cite{defossez2022encodec}, DAC \cite{kumar2023dac}, and WavTokenizer \cite{ji2024wavtokenizer}, as well as codec-based speech generation systems such as VALL-E \cite{wang2023valle}, CosyVoice \cite{du2024cosyvoice}, and Moshi \cite{defossez2024moshi}, reconstruct speech through quantized acoustic representations, preserving content and speaker identity while discarding details that are not stable under codec reconstruction.  This makes latent-side audio watermarking relevant, because the watermark can be designed around the representations that neural codecs preserve.  Nevertheless, watermark evidence placed even in generation-specific latent variables can still be weakened after resynthesis.  Studies have shown that the performance of previous watermarking methods degrades under neural codec resynthesis \cite{wu2025comparative}.

This motivates audio watermarking methods that are robust to neural-codec resynthesis attacks.  As one solution, we propose NeuMark, a codec-latent watermarking framework designed to improve robustness against neural-codec resynthesis.  We use SpeechTokenizer \cite{zhang2023speechtokenizer} because its residual vector quantization (RVQ) structure provides multiple codec-aligned acoustic token streams, allowing watermark evidence to be embedded in representations targeted by neural-codec reconstruction.  NeuMark embeds a recoverable message directly into codec-aligned latent representations and decodes the modified latent sequence into watermarked speech.  During training, NeuMark simulates DSP attacks and neural-codec resynthesis, and is trained to perform both frame-level watermark detection and message recovery.  Because latent-space watermarking is bounded by the reconstruction quality of the codec, we also evaluate transparency against both the original waveform and a codec-reconstructed reference.

The main contributions of this paper are as follows:
\begin{itemize}
  \item We propose NeuMark, a codec-latent audio watermarking method designed for robustness against neural-codec resynthesis, and compare it with waveform, timbre-feature, and latent/generation-aware baselines under DSP edits and codec-resynthesis attacks.
  \item We reveal a robustness--transparency trade-off specific to latent codec watermarking through reconstruction-referenced and original-referenced training variants.  Our evaluation against DSP edits and neural-codec resynthesis shows that codec-aligned embedding substantially improves robustness to resynthesis, while metrics using reconstructed and original references separate watermark distortion from codec reconstruction effects.
\end{itemize}

The code is publicly available at \url{https://github.com/goannan/NeuMark}.

\section{Comparison with Latent-Side Methods}
TraceableSpeech \cite{zhou2024traceablespeech}, VoiceMark \cite{li2025voicemark}, and NeuMark all embed watermark evidence in latent representations, but they differ in where the message is injected, what attacks are simulated during training, and what reference is used for quality optimization, as summarized in Table~\ref{tab:latent_methods}.  TraceableSpeech injects the watermark into the representation after RVQ-layer aggregation before decoding.  VoiceMark uses SpeechTokenizer, keeps the first VQ layer as content, and embeds into the speaker-specific VQ 2--8 layers for zero-shot voice-cloning transfer.  NeuMark instead injects the message into each RVQ layer separately, so the watermark is distributed over all codec layers for codec-resynthesis robustness.  For simulated attacks during training, TraceableSpeech uses DSP and EnCodec augmentation, VoiceMark uses DSP and voice-clone simulated augmentation, and NeuMark uses DSP, masking, and EnCodec augmentation.

\begin{table*}[t]
\centering
\scriptsize
\renewcommand{\arraystretch}{1.08}
\caption{Comparison of latent-side or generation-aware audio watermarking methods.}
\label{tab:latent_methods}
\begin{tabular*}{\textwidth}{@{\extracolsep{\fill}}>{\raggedright\arraybackslash}p{0.16\textwidth}>{\raggedright\arraybackslash}p{0.21\textwidth}>{\raggedright\arraybackslash}p{0.31\textwidth}>{\raggedright\arraybackslash}p{0.16\textwidth}@{}}
\toprule
Method & Embed position & Simulated attacks & Reference \\
\midrule
TraceableSpeech & After RVQ aggregation & DSP and EnCodec & Original \\
VoiceMark & VQ 2--8 speaker layers & DSP and voice-clone simulated augmentation & Original \\
NeuMark (ours) & Each RVQ layer & DSP, masking, and EnCodec & Original or codec recon. \\
\bottomrule
\end{tabular*}
\end{table*}

\section{Proposed Method}
\subsection{Overview}
\begin{figure*}[t]
\centering
\includegraphics[width=0.94\textwidth]{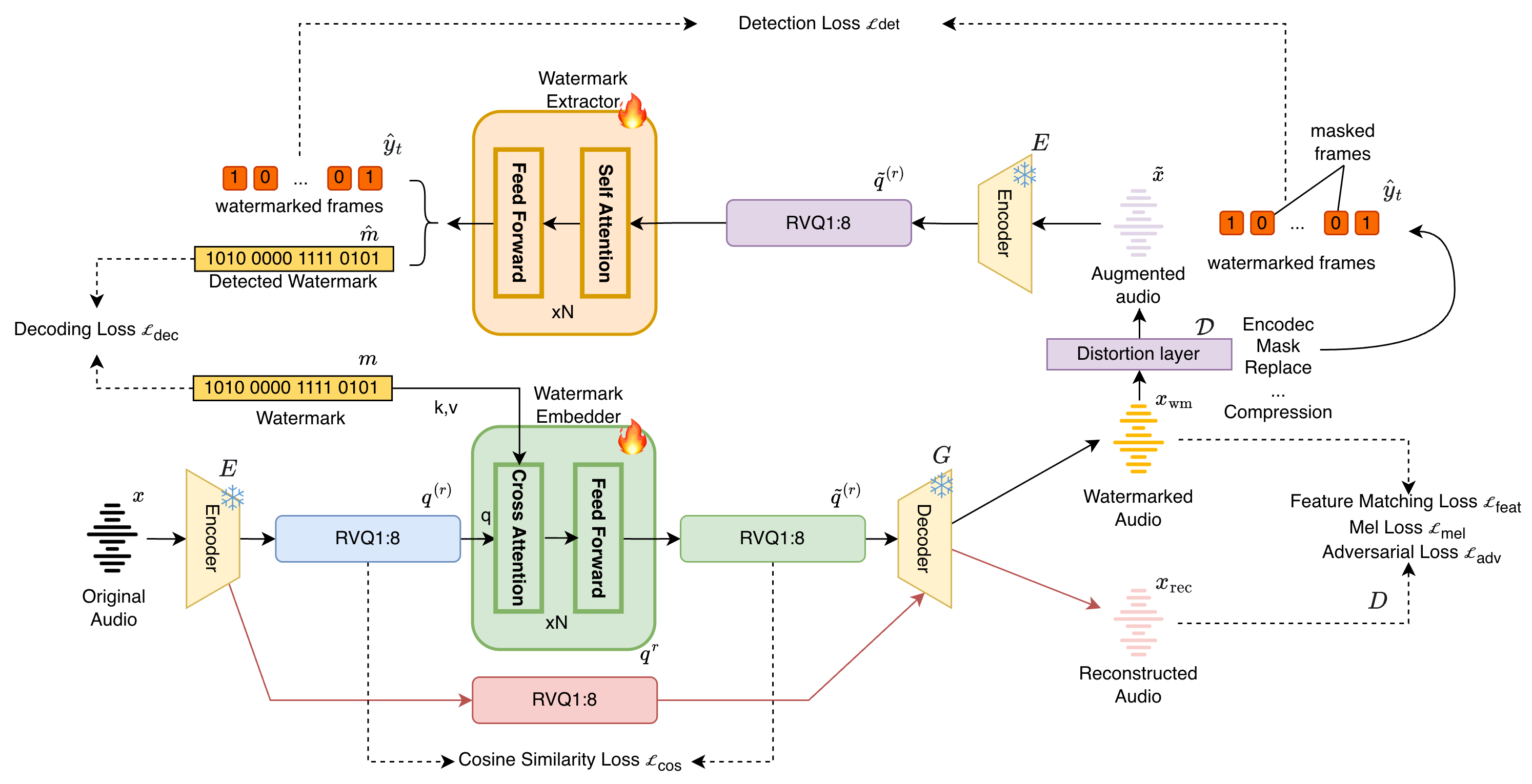}
\caption{The overall architecture of our proposed NeuMark.}
\label{fig:pipeline}
\end{figure*}

Fig. \ref{fig:pipeline} summarizes the NeuMark training pipeline.  Given an input utterance $x\in\mathbb{R}^{L}$ and a 16-bit message $m\in\{0,1\}^{16}$, where $L$ is the waveform length, the frozen SpeechTokenizer encoder $E$ produces $e\in\mathbb{R}^{1024\times T}$.  Here, 1024 denotes the SpeechTokenizer latent channel dimension, and $T$ is the number of latent frames.  Its eight RVQ streams are $q^{(r)}\in\mathbb{R}^{1024\times T}$ for $r=1,\ldots,8$, with $T=L/320$ (e.g., $T=150$ for a 3-second 16 kHz segment).  NeuMark modifies these RVQ streams with a trainable Transformer-based watermark encoder and decodes the result into watermarked speech $\xwm$ with the frozen SpeechTokenizer decoder $G$.  During training, $\xwm$ is passed through a stochastic distortion operator $\mathcal{D}$ before a trainable extractor predicts frame-level watermark probabilities and message bits.

\subsection{Cross-Attention-Based Watermark Encoder}
The 16-bit message $m$ is converted into a sequence of learnable message-memory vectors, which condition each RVQ stream through a Transformer cross-attention block \cite{vaswani2017attention}.  The cross-attended features are projected back to the SpeechTokenizer latent dimension and added to each RVQ stream through a residual watermark encoder.  The modified RVQ streams are then summed and decoded by the frozen SpeechTokenizer decoder $G$ to obtain $\xwm$.  The pretrained tokenizer remains fixed throughout training.

\subsection{Training Distortion Operator}
Training uses a stochastic distortion operator $\mathcal{D}$ applied to $\xwm$.  For each mini-batch, $\mathcal{D}$ samples one augmentation from the training distortion pool and produces the distorted watermarked speech $\tilde{x}=\mathcal{D}(\xwm)$.  The training pool and sampling weights are described in Section~\ref{sec:training_setup}; masking is used only as a training augmentation, not as an evaluation attack.

\subsection{Training Reference Variants}
\label{sec:reference_variants}
NeuMark computes a reference reconstruction $\xrec = G(E(x))$, where $E$ and $G$ are the frozen SpeechTokenizer encoder and decoder.  This reference is important because latent-space watermarking cannot exceed the reconstruction fidelity of the codec itself.  We define two NeuMark training variants that differ only in the choice of the reference target $\xref$ used across all quality losses and discriminator training:
\begin{itemize}
  \item \textbf{NeuMark R-R (Reconstruction Reference):} Sets $\xref=\xrec$, using the codec-reconstructed audio as the reference target.  This isolates watermark-induced distortion from the distortion already introduced by the codec.
  \item \textbf{NeuMark R-O (Original Reference):} Sets $\xref=x$, using the original source audio as the reference target.  This trains the embedder to both embed the watermark and compensate for the codec reconstruction error.
\end{itemize}

\subsection{Training Objective}
The training objective of NeuMark combines quality-preserving losses (adversarial loss $\mathcal{L}_{\mathrm{adv}}$, feature matching loss $\mathcal{L}_{\mathrm{feat}}$, mel-spectral reconstruction loss $\mathcal{L}_{\mathrm{mel}}$, and cosine similarity loss $\mathcal{L}_{\mathrm{cos}}$) and watermark-extraction losses (message decoding loss $\mathcal{L}_{\mathrm{dec}}$ and watermark presence detection loss $\mathcal{L}_{\mathrm{det}}$).

The adversarial loss encourages watermarked speech to match the distribution of natural and codec-reconstructed speech:
\begin{equation}
\mathcal{L}_{\mathrm{adv}} =
\mathbb{E}\left[(D(\xwm)-1)^2\right],
\end{equation}
where $D(\cdot)$ is the scalar output of the least-squares adversarial discriminator and $\mathbb{E}[\cdot]$ denotes expectation over training samples and sampled augmentations.  The discriminator is trained to classify the reference audio $\xref$ as real and the watermarked audio $\xwm$ as fake:
\begin{equation}
\mathcal{L}_{\mathrm{D}} =
\mathbb{E}\left[(D(\xref)-1)^2\right] + \mathbb{E}\left[D(\xwm)^2\right].
\end{equation}

In these equations, $\xref$ denotes the quality reference target, which is either the original speech $x$ or the codec reconstruction $\xrec=G(E(x))$ depending on the training variant described in Section~\ref{sec:reference_variants}.

The feature matching loss is shown as follows:
\begin{equation}
\mathcal{L}_{\mathrm{feat}} =
\sum_{i=1}^{J}\frac{1}{N_i}
\left\|D_i(\xref)-D_i(\xwm)\right\|_1,
\end{equation}
where $J$ is the number of discriminator feature layers used for matching, $D_i(\cdot)$ is the feature map from the $i$-th discriminator layer, and $N_i$ is the number of scalar elements in that feature map.

The multi-scale mel-spectrogram loss is
\begin{equation}
\mathcal{L}_{\mathrm{mel}} =
\sum_{s\in \mathcal{S}}
\left\|\mathrm{Mel}_{s}(\xref)-\mathrm{Mel}_{s}(\xwm)\right\|_1 .
\end{equation}
Here, $\mathcal{S}$ is the set of mel-spectrogram scales, and $\mathrm{Mel}_{s}(\cdot)$ denotes the mel-spectrogram computed with scale setting $s$.

The latent cosine loss constrains the direction of the content latent before and after embedding:
\begin{equation}
\mathcal{L}_{\mathrm{cos}} =
1 -
\frac{z_c^\top z'_c}{\|z_c\|_2\|z'_c\|_2},
\end{equation}
where $z_c$ and $z'_c$ denote flattened content-latent representations before and after watermark embedding, respectively.

The extractor encodes distorted watermarked speech $\tilde{x}$ with the frozen SpeechTokenizer encoder and predicts a 16-bit message estimate $\mhat\in[0,1]^{16}$ and frame-level watermark probabilities $\hat{\mathbf{y}}\in[0,1]^{T}$.  The message decoding loss is
\begin{equation}
\mathcal{L}_{\mathrm{dec}} =
\mathrm{CE}(m,\mhat),
\end{equation}
where $\mathrm{CE}(\cdot,\cdot)$ denotes binary cross entropy over the 16 payload bits.  The extractor also outputs logits $l\in\mathbb{R}^{T}$ with $\hat{y}_{t}=\sigma(l_{t})$, where $\sigma(\cdot)$ is the sigmoid function.  The detection loss is frame-level binary cross entropy with a VAD speech mask:
\begin{equation}
\mathcal{L}_{\mathrm{det}} =
-\frac{1}{V+\epsilon}\sum_{t=1}^{T}v_t
\left[
y_t\log\hat{y}_t+(1-y_t)\log(1-\hat{y}_t)
\right].
\end{equation}
Here, for notational simplicity the batch index is omitted: $y_t\in\{0,1\}$ is the target watermark-presence label at frame $t$ ($1$ for watermarked speech and $0$ for clean or masked negative regions), $v_t\in\{0,1\}$ is the VAD label indicating whether frame $t$ contains speech, $V=\sum_{t=1}^{T}v_t$ is the number of speech frames, and $\epsilon$ prevents division by zero.  VAD labels are computed on the clean training utterance, aligned to the extractor frame rate, and applied only as a loss mask so that detection errors in silent frames are not emphasized.  Positive samples use distorted watermarked speech, while negative samples use clean speech or masked regions.

The final objective for the watermark encoder and extractor is the weighted sum
\begin{equation}
\begin{aligned}
\mathcal{L}_{\mathrm{total}} ={}&
\lambda_{\mathrm{adv}}\mathcal{L}_{\mathrm{adv}}
+ \lambda_{\mathrm{feat}}\mathcal{L}_{\mathrm{feat}}
+ \lambda_{\mathrm{mel}}\mathcal{L}_{\mathrm{mel}} \\
&+ \lambda_{\mathrm{cos}}\mathcal{L}_{\mathrm{cos}}
+ \lambda_{\mathrm{dec}}\mathcal{L}_{\mathrm{dec}}
+ \lambda_{\mathrm{det}}\mathcal{L}_{\mathrm{det}} .
\end{aligned}
\end{equation}
The discriminator is optimized separately with $\mathcal{L}_{\mathrm{D}}$.

\section{Experimental Setup}
\subsection{Data}
All watermark models were evaluated with the same test set: LibriSpeech test-clean \cite{panayotov2015librispeech}, containing 2620 utterances.  NeuMark was trained on the LibriTTS train-clean-100, train-clean-360, and train-other-500 subsets at 24 kHz \cite{zen2019libritts}.

\subsection{Baselines}
Using 16-bit payloads throughout, we compared NeuMark with five baselines: WavMark \cite{chen2023wavmark},\footnote{\url{https://github.com/wavmark/wavmark}.} AudioSeal \cite{sanroman2024audioseal},\footnote{\url{https://github.com/facebookresearch/audioseal}.} Timbre Watermarking \cite{liu2023timbrewm},\footnote{\url{https://github.com/TimbreWatermarking/TimbreWatermarking}.} TraceableSpeech \cite{zhou2024traceablespeech},\footnote{\url{https://github.com/zjzser/TraceableSpeech}.} and VoiceMark \cite{li2025voicemark}.\footnote{\url{https://huggingface.co/spaces/haiyunli/VoiceMark}.}  TraceableSpeech provided code but no pretrained checkpoint, so we retrained it on LibriTTS with EnCodec resynthesis augmentation.  In the quality evaluation (Table~\ref{tab:quality}), WavMark, AudioSeal, and Timbre Watermarking were treated as non-reconstruction baselines because they lack paired codec-reconstruction references.

\subsection{Training Details}
\label{sec:training_setup}
The default weights for the loss terms were $\lambda_{\mathrm{adv}}=1$, $\lambda_{\mathrm{feat}}=2$, $\lambda_{\mathrm{mel}}=2$, $\lambda_{\mathrm{cos}}=2$, $\lambda_{\mathrm{dec}}=10$, and $\lambda_{\mathrm{det}}=1$.  Each NeuMark variant was trained for 150,000 steps with a global batch size of 32 on four NVIDIA GeForce RTX 3090 GPUs, requiring approximately 97 hours.  The training distortion pool included an identity path, DSP distortions broadly following the AudioSeal-style DSP edits, EnCodec neural codec resynthesis at 3, 6, and 12 kbps, and masking.  The masking augmentation randomly selected 20\% of the time samples in each watermarked waveform.  The selected samples were directly modified on the waveform time axis: for silence masking, they were set to zero; for original-audio replacement, they were replaced by the temporally aligned samples from the original utterance $x$, while all unselected samples were left unchanged.  The sampling weights were set to 10 for identity, 1 for each DSP distortion, 1 for each neural codec distortion, and 5 for masking.

\subsection{Evaluation Attacks}
The test-time robustness benchmark included the DSP edits None, Bandpass, Echo, White, Highpass, Lowpass, Pink, Resample, Smooth, Fast, Boost, and Duck, following the naming used in AudioSeal.\footnote{\url{https://github.com/facebookresearch/audiocraft}.}  Fast denoted speed perturbation from 0.8$\times$ to 1.2$\times$; Boost and Duck denoted volume scaling by $+10\%$ and $-10\%$, respectively.  Neural resynthesis attacks included DAC at 16 kHz and 24 kHz, EnCodec at 12, 6, and 3 kbps, and WavTokenizer.  We used EnCodec as the seen neural-codec attack because the public pretrained baseline models used in this comparison were generally trained with EnCodec-style codec augmentation, allowing us to evaluate released models without retraining all baselines.  DAC and WavTokenizer were therefore treated as unseen attacks to test generalization beyond the codec used during robustness training.

\subsection{Metrics}
For robustness, we evaluated the watermark detection rate ($\Det$) and bit recovery rate ($\Bit$).  Frame-level detection probabilities were averaged into utterance-level scores $p_i$.
Detection was computed as balanced soft accuracy over paired watermarked and clean utterances:
\begin{equation}
\Det = \frac{1}{2}\left( \frac{1}{N}\sum_{i=1}^{N}p_{i,\mathrm{wm}} + \frac{1}{N}\sum_{i=1}^{N}(1-p_{i,\mathrm{clean}}) \right).
\end{equation}
Bit recovery measured the fraction of correctly recovered payload bits:
\begin{equation}
\Bit = \frac{1}{NK}\sum_{i=1}^{N}\sum_{k=1}^{K}\mathbf{1}\left[\hat{m}_{i,k}=m_{i,k}\right],
\end{equation}
where $m_{i,k}$ and $\hat{m}_{i,k}$ are the target and recovered value of bit $k$ for utterance $i$, respectively.

Perceptual quality was measured using PESQ \cite{pesq}, STOI \cite{taal2011stoi}, and SI-SNR \cite{leroux2019sdr} under the R-O (Original Reference) and R-R (Reconstruction Reference) protocols introduced in Section~\ref{sec:reference_variants}.  R-O metrics compared watermarked speech against the original audio $x$, while R-R metrics compared it against the reconstructed control $\xrec$.  R-R evaluation was not applicable to non-reconstruction baselines.

\section{Results and Analysis}
\begin{table*}[t]
\centering
{\footnotesize
\setlength{\tabcolsep}{2pt}
\renewcommand{\arraystretch}{1.05}
\setlength{\abovecaptionskip}{1pt}
  \begin{threeparttable}
  \caption{Robustness under signal-processing and neural-codec resynthesis attacks.  Each cell reports Det/Bit.}
  \label{tab:robustness}
  \begin{tabular*}{\textwidth}{@{\extracolsep{\fill}}lccccccc}
  \toprule
  Attack & WavMark & AudioSeal & TimbreWM & TraceableSpeech & VoiceMark & NeuMark R-R & NeuMark R-O \\
  \midrule
  None & \textbf{1.00} / \textbf{1.00} & \textbf{1.00} / \textbf{1.00} & 0.70 / \textbf{1.00} & 0.78 / \textbf{1.00} & 0.83 / 0.99 & 0.99 / \textbf{1.00} & \textbf{1.00} / \textbf{1.00} \\
  Bandpass & \textbf{1.00} / \textbf{1.00} & \textbf{1.00} / \textbf{1.00} & 0.69 / \textbf{1.00} & 0.80 / \textbf{1.00} & 0.84 / 0.99 & 0.99 / \textbf{1.00} & \textbf{1.00} / \textbf{1.00} \\
  Echo & \textbf{0.98} / 0.94 & 0.82 / 0.81 & 0.67 / \textbf{1.00} & 0.77 / \textbf{1.00} & 0.71 / 0.98 & 0.93 / 0.97 & 0.95 / 0.93 \\
  White & 0.98 / 0.97 & 0.98 / \textbf{1.00} & 0.63 / \textbf{1.00} & 0.79 / \textbf{1.00} & 0.77 / 0.96 & 0.96 / 0.98 & \textbf{0.99} / \textbf{1.00} \\
  Highpass & \textbf{1.00} / \textbf{1.00} & \textbf{1.00} / \textbf{1.00} & 0.70 / \textbf{1.00} & 0.79 / \textbf{1.00} & 0.83 / 0.99 & 0.99 / 0.99 & \textbf{1.00} / \textbf{1.00} \\
  Lowpass & \textbf{1.00} / \textbf{1.00} & \textbf{1.00} / \textbf{1.00} & 0.67 / \textbf{1.00} & 0.79 / 0.96 & 0.78 / 0.95 & 0.98 / 0.99 & \textbf{1.00} / \textbf{1.00} \\
  Pink & \textbf{1.00} / \textbf{1.00} & \textbf{1.00} / \textbf{1.00} & 0.70 / \textbf{1.00} & 0.78 / \textbf{1.00} & 0.81 / 0.99 & 0.99 / \textbf{1.00} & \textbf{1.00} / \textbf{1.00} \\
  Resample & \textbf{1.00} / \textbf{1.00} & \textbf{1.00} / \textbf{1.00} & 0.70 / \textbf{1.00} & 0.78 / \textbf{1.00} & 0.83 / 0.97 & 0.99 / \textbf{1.00} & \textbf{1.00} / \textbf{1.00} \\
  Smooth & \textbf{1.00} / 0.99 & 0.93 / 0.76 & 0.66 / \textbf{1.00} & 0.77 / 0.96 & 0.69 / 0.74 & 0.94 / 0.96 & 0.98 / 0.99 \\
  Fast (0.8$\times$--1.2$\times$) & 0.51 / 0.51 & 0.51 / 0.52 & 0.59 / 0.49 & 0.65 / 0.55 & 0.75 / 0.62 & 0.88 / 0.76 & \textbf{0.90} / \textbf{0.82} \\
  Boost & \textbf{1.00} / \textbf{1.00} & \textbf{1.00} / \textbf{1.00} & 0.70 / \textbf{1.00} & 0.78 / \textbf{1.00} & 0.84 / 0.99 & 0.99 / \textbf{1.00} & \textbf{1.00} / \textbf{1.00} \\
  Duck & \textbf{1.00} / \textbf{1.00} & \textbf{1.00} / \textbf{1.00} & 0.70 / \textbf{1.00} & 0.78 / \textbf{1.00} & 0.82 / 0.99 & 0.99 / \textbf{1.00} & \textbf{1.00} / \textbf{1.00} \\
  DSP Avg. & 0.95 / 0.95 & 0.93 / 0.92 & 0.67 / 0.95 & 0.77 / 0.95 & 0.79 / 0.92 & 0.97 / 0.97 & \textbf{0.98} / \textbf{0.98} \\
  \midrule
  EnCodec 12 kbps & 0.50 / 0.50 & 0.84 / 0.88 & 0.51 / 0.67 & 0.64 / 0.82 & 0.77 / 0.96 & 0.92 / 0.96 & \textbf{0.98} / \textbf{0.99} \\
  EnCodec 6 kbps & 0.50 / 0.50 & 0.68 / 0.61 & 0.50 / 0.57 & 0.58 / 0.72 & 0.76 / 0.95 & 0.88 / 0.93 & \textbf{0.95} / \textbf{0.98} \\
  EnCodec 3 kbps & 0.50 / 0.50 & 0.61 / 0.53 & 0.50 / 0.53 & 0.54 / 0.61 & 0.74 / 0.90 & 0.79 / 0.84 & \textbf{0.89} / \textbf{0.92} \\
  DAC 16 kHz* & 0.50 / 0.50 & 0.60 / 0.31 & 0.52 / 0.83 & 0.62 / 0.75 & 0.82 / 0.97 & 0.95 / 0.98 & \textbf{0.98} / \textbf{1.00} \\
  DAC 24 kHz* & 0.55 / 0.91 & 0.98 / 0.85 & 0.60 / 0.99 & 0.76 / 0.97 & 0.82 / 0.98 & 0.98 / 0.99 & \textbf{1.00} / \textbf{1.00} \\
	  WavTokenizer* & 0.50 / 0.50 & 0.50 / 0.50 & 0.50 / 0.50 & 0.49 / 0.50 & \textbf{0.62} / \textbf{0.55} & 0.51 / 0.53 & 0.50 / 0.54 \\
	  Neural Codec Avg. & 0.51 / 0.57 & 0.70 / 0.61 & 0.52 / 0.68 & 0.60 / 0.73 & 0.75 / 0.88 & 0.84 / 0.87 & \textbf{0.88} / \textbf{0.91} \\
	  \midrule
	  Average & 0.81 / 0.82 & 0.86 / 0.82 & 0.62 / 0.87 & 0.72 / 0.88 & 0.78 / 0.92 & 0.93 / 0.94 & \textbf{0.95} / \textbf{0.95} \\
	  \bottomrule
  \end{tabular*}
  \begin{tablenotes}[flushleft]
	  \item[*] Unseen attacks during training. DSP Avg. and Neural Codec Avg. are computed uniformly over their 12 and 6 constituent attacks, respectively; Average is computed uniformly over all 18 attacks.
	  \end{tablenotes}
	  \end{threeparttable}
  \vspace{0.2em}

	  \begin{threeparttable}
	  \caption{Perceptual quality of watermarked speech.  In the metric columns, R-O and R-R denote the evaluation reference audio: original audio $x$ and reconstructed control $\xrec$, respectively.  In the NeuMark model names, R-O and R-R denote the training reference target $\xref=x$ or $\xref=\xrec$.}
	  \label{tab:quality}
  \begin{tabular*}{\textwidth}{@{\extracolsep{\fill}}lcccccc}
  \toprule
  Method & PESQ R-O & STOI R-O & SI-SNR R-O & PESQ R-R & STOI R-R & SI-SNR R-R \\
  \midrule
  WavMark & 4.233 & 0.997 & 37.298 & -- & -- & -- \\
  AudioSeal & 4.429 & 0.998 & 26.828 & -- & -- & -- \\
  TimbreWM & 3.862 & 0.993 & 27.949 & -- & -- & -- \\
  \midrule
  TraceableSpeech* & 2.495 & \textbf{0.927} & 1.055 & 4.236 & \textbf{0.989} & 13.265 \\
  VoiceMark* & 2.136 & 0.910 & \textbf{1.176} & 3.320 & 0.970 & 10.366 \\
  NeuMark R-R* & \textbf{2.501} & 0.920 & 0.839 & \textbf{4.356} & \textbf{0.989} & \textbf{13.934} \\
  NeuMark R-O* & 2.318 & 0.907 & -0.191 & 3.349 & 0.956 & 5.754 \\
  \bottomrule
  \end{tabular*}
  \begin{tablenotes}[flushleft]
  \item[*] Codec-latent methods.
	  \end{tablenotes}
	  \end{threeparttable}
}
\end{table*}

\subsection{Watermark Extraction Robustness}
As summarized in Table~\ref{tab:robustness}, NeuMark R-O achieves the highest average robustness, followed closely by NeuMark R-R.  Under DSP attacks, waveform-domain methods such as WavMark and AudioSeal remain highly competitive.  However, they degrade severely under neural-codec resynthesis, where NeuMark's alignment with SpeechTokenizer latents preserves higher detection and bit recovery.  The neural-codec average highlights this difference, and the unseen DAC and WavTokenizer results further test generalization beyond the codec used during robustness training.  Speed perturbation remains challenging for all methods because it changes temporal alignment, but the NeuMark variants still yield the highest extraction accuracy.  WavTokenizer's extreme compression strongly degrades watermarking capacity, suggesting that future watermark designs need stronger invariance to single-quantizer resynthesis rather than relying only on multi-RVQ residual capacity.

\subsection{Perceptual Quality}
As shown in Table~\ref{tab:quality}, non-reconstruction baselines (e.g., AudioSeal) yield higher fidelity against the original signal because they bypass codec reconstruction.  For latent-space methods, R-R evaluation is more informative since it isolates watermark-only distortion.

Under R-R evaluation, NeuMark R-R achieves the highest PESQ and SI-SNR and matches TraceableSpeech in STOI at the reported precision.  NeuMark R-O improves robustness but has lower R-R quality.  This trade-off occurs because training against the original audio forces the R-O model to perform both embedding and codec restoration, introducing larger acoustic perturbations.

\section{Conclusion}
This paper introduced NeuMark, a latent-space audio watermarking framework for robust detection and 16-bit message recovery under neural-codec resynthesis.  NeuMark uses a Transformer cross-attention-based watermark encoder over SpeechTokenizer latents and is trained with a multi-objective loss that jointly preserves speech quality and improves extraction robustness.  The proposed evaluation protocol compares audio-domain, feature-/timbre-domain, and latent/generation-aware watermarks under DSP and neural codec attacks, with quality measured against both original and reconstruction references.

The current evaluation treats the embedder and extractor as a watermarking module around a pretrained tokenizer.  A full TTS integration, where a codec language model directly generates watermarked acoustic tokens, remains future work.  The most difficult attack is currently WavTokenizer-style extreme compression, where one quantizer must represent the essential speech signal and leaves little room for watermark evidence.  Future work will explore stronger temporal redundancy, watermark objectives aligned with single-quantizer invariants, and multilingual/noisy-speech evaluation.

\section*{Acknowledgment}
This work was supported in part by JSPS KAKENHI Grant Number 26H02530, and in part by the BRIDGE Program (R7-H05), implemented by the Cabinet Office, Government of Japan.

\printbibliography

\end{document}